\documentclass [prb,twocolumn,superscriptaddress] {revtex4}
\usepackage[]{amssymb,amsmath,amsfonts}
\usepackage[]{graphicx}

\begin{document}
\title{Design of a surface acoustic wave mass sensor in the 100 GHz range}

\author{Damiano Nardi} \email[]{damiano.nardi@jila.colorado.edu}
\affiliation{JILA, University of Colorado at Boulder, Boulder, Colorado 80309, United States}
\affiliation{i-LAMP and Dipartimento di Matematica e Fisica, Universit\`a Cattolica del Sacro Cuore, I-25121 Brescia, Italy}

\author{Elisa Zagato}
\author{Gabriele Ferrini}
\author{Claudio Giannetti}
\author{Francesco Banfi}
\affiliation{i-LAMP and Dipartimento di Matematica e Fisica, Universit\`a Cattolica del Sacro Cuore, I-25121 Brescia, Italy}

\date{\today}

\begin{abstract}
A design for photoacoustic mass sensors operating above 100 GHz is proposed. The design is based on impulsive optical excitation of a pseudosurface acoustic wave in a surface phononic crystal with nanometric periodic grating, and on time-resolved extreme ultraviolet detection of the pseudosurface acoustic wave frequency shift upon mass loading the device. The present design opens the path to sensors operating in a frequency range currently unaccessible to electro-acoustical transducers, providing enhanced sensitivity, miniaturization and incorporating time-resolving capability while forgoing the piezoelectric substrate requirement.
\end{abstract}

\maketitle

\indent The field of microscale mass sensors has been booming recently, driven by the advances in nanopatterning techniques and the increasing request for devices capable of minute amounts of matter detection, most notably for biological and environmental interests. A variety of sensors have been implemented, among them surface acoustic waves (SAW)-based devices.\cite{Gronewold2007, Voinova2009}\\
\indent A SAW is an elastic wave that propagates confined to the surface of a semi-infinite medium, the penetration depth being a fraction of its wavelength $\lambda=v_{SAW}/\nu$, with $v_{SAW}$ and $\nu$ as the SAW velocity and frequency, respectively. The interaction with any medium in contact with the surface affects frequency and lifetime of the wave itself. As the SAW frequency $\nu$ increases, so do its surface confinement ($\sim$$\lambda^{-1}$) and the device's sensitivity to mass loading.\\
\indent In typical SAW-based devices, SAWs are launched and detected via two interdigital transducers (IDT) patterned on a piezoelectric substrate. Upon mass loading the free surface between IDTs, the unperturbed SAW frequency $\nu_{0}$ shifts downward of an amount $\Delta\nu$. Measurement of $\Delta\nu/\nu_{0}$ enables quantification of the bound mass.\cite{Vellekoop1998} These devices perform well in terms of $\Delta\nu$, but the drawbacks are the maximum operating frequency bounded to the GHz range due to speed limits in the electronics, the lack of fast temporal resolution, and the piezoelectric substrate requirement.\\
\indent A strategy is here outlined to overcome these limitations. It relies on all-optical generation of a pseudosurface acoustic wave (pseudo-SAW) in a hypersonic surface phononic crystal (SPC),\cite{Giannetti2009, Nardi2011, Maris1998} and on time-resolved extreme ultraviolet (EUV) detection of the pseudo-SAW frequency in a diffractive scheme.\cite{Tobey2004, Siemens2009} The device we propose relies on a SPC made of periodic Al stripes (Ni stripes are also investigated) deposited on sapphire (stripes' periodicity $p=50$~nm, height $h=2$~nm, filling fraction $f=d/p=0.2$, mass density $\rho_{Al}=2700$~Kg/m$^{3}$). The stripes' width $d\sim10$~nm is within reach of state of the art e-beam lithography. The pseudo-SAW frequency $\nu_0$, set by $v_{SAW}/p$, is expected to shift of $\Delta\nu$ upon mass loading the SPC,\cite{Nardi2009} enabling mass detection. The present scheme gives access to operating acoustic frequencies beyond 100 GHz, enhancing the device's sensitivity, incorporates ultrafast time-resolving capabilities, forgoes the piezoelectric substrate requirement, and allows for increased miniaturization with respect to standard mass-sensor technology.\\
\begin{figure}[b]
\centering
\includegraphics[bb=98 88 508 434,keepaspectratio,clip,width=0.7\columnwidth]{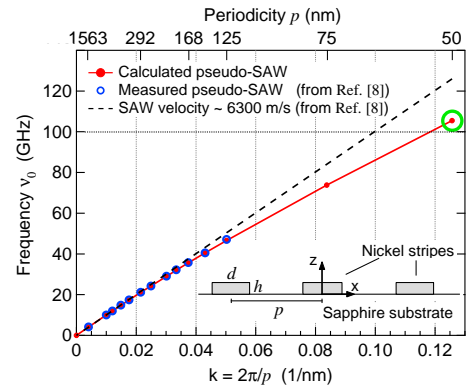}
\caption{(Color online) Pseudo-SAW frequency vs. wavevector dispersion for Ni-based SPCs (inset), with $h=10$ nm, $f=0.5$ and $p$ as indicated on top $x$-axis. The reported substrate's Rayleigh wave linear dispersion (black dashed line) and experimental data (blue empty circles) are taken from Siemens et al., see Ref.~[\onlinecite{Siemens2009}]. The pseudo-SAW frequencies (red circles) are calculated for the same SPC configurations experimentally investigated in Ref.~[\onlinecite{Siemens2009}]. The pseudo-SAW dispersion is calculated beyond a perturbative approach (red line), see Ref.~[\onlinecite{Nardi2009}]. The green circle highlights the pseudo-SAW frequency calculated for $p=50$~nm.}
\label{Fig1}
\end{figure}
\begin{figure*}[t]
\centering
\includegraphics[bb=100 12 896 440,keepaspectratio,clip,width=1.7\columnwidth]{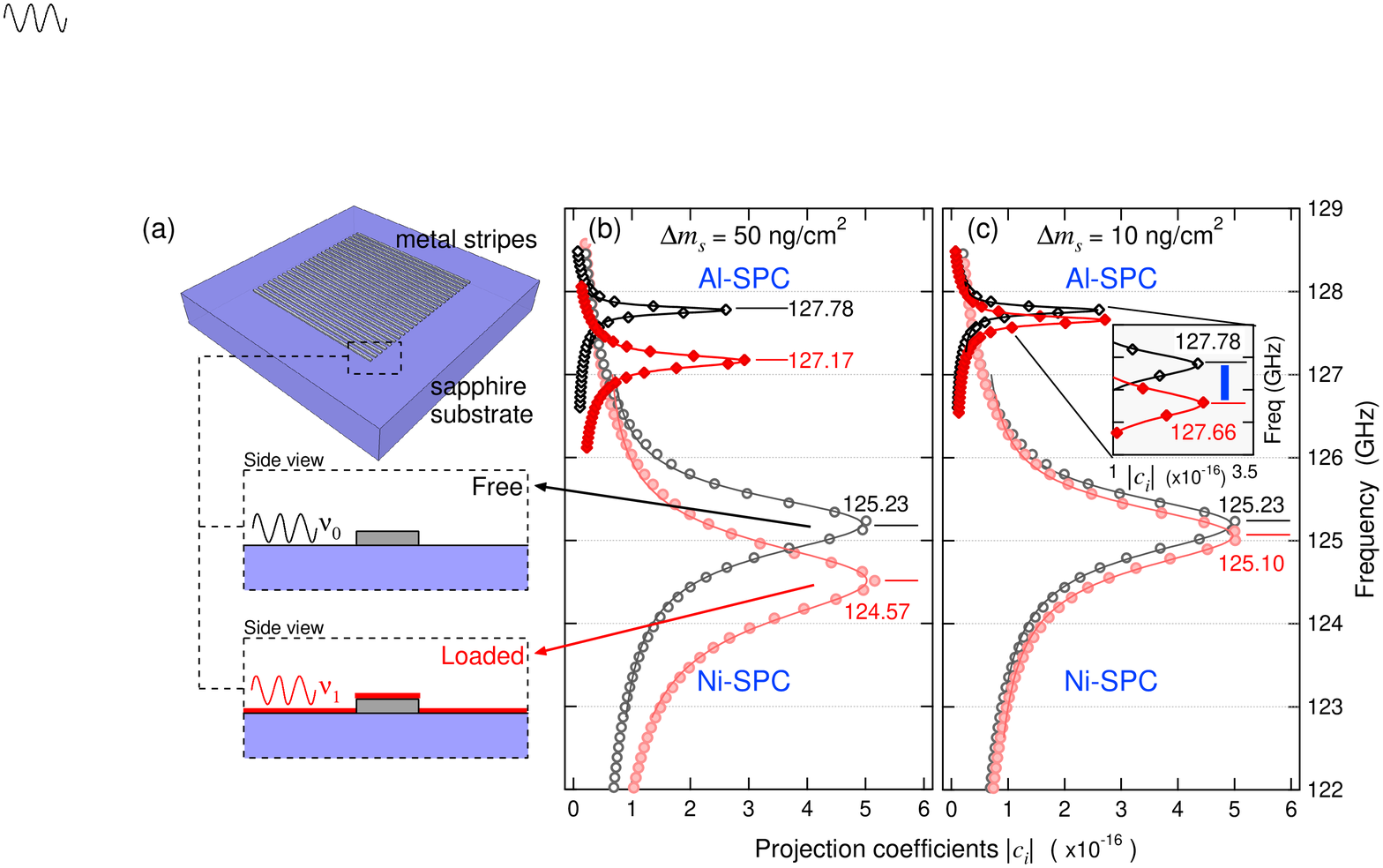}
\caption{(Color on line) (a) SPC mass sensor: schematic drawing of free (top) and loaded (bottom) device. The absolute value of the projection coefficients $\left|c_i\right|$ vs. frequency, calculated for free (empty marks) and loaded (filled marks) device, are reported for both Al-SPC (diamonds) and Ni-SPC (circles). The data sets are fit with a Fano line shape profile. Two mass loading configurations are investigated: (b) $50$~ng/cm$^{2}$, and (c) $10$ ng/cm$^{2}$. Detailed view of the frequency shift of Al-SPC upon mass loading is reported in the inset. The blue bar is the $\textit{extrinsic}$ frequency resolution $\Delta\nu_{ext}=0.1$~GHz.}
\label{Fig2}
\end{figure*}
\indent The SPC is illuminated by an ultrafast IR laser pump pulse.\cite{note2} Upon energy absorption, the stripes' temperature increase triggers a spatially modulated thermal expansion.\cite{Banfi2010} The main contribution to the resulting displacement field $\textbf{u}\left(\textbf{r},t\right)$, once projected over the set of eigenmodes $\left\{\textbf{u}_{i}(\textbf{r})\right\}$ of the phononic crystal, solutions of the acoustic eigenvalue problem, is found to be a symmetric pseudo-SAW $\textbf{u}_{SAW}$ with wavelength $\lambda=p$,\cite{Nardi2011} and frequency $\nu_0$. This solution corresponds to the first harmonic at the center of the surface Brillouin zone (Bloch wavevector $k_{x}=0$ and band index $n$=1). Spatially, the pseudo-SAW is partially localized on the nanostructures and radiates elastic energy into the bulk.\cite{Nardi2009}\\
\indent The generated pseudo-SAW acoustically modulates the SPC, which serves as an optical $\textit{phase}$ diffraction grating for a second time-delayed EUV laser pulse with wavelength $\lambda_{probe}\sim p$. The EUV pulse hence probes the frequency $\nu_{0}$ of acoustic modulation. The relative variation in the time-domain of the diffracted probe signal $\Delta I_{1d}/I_{1d}$, as measured on the first order of diffraction spot\cite{note1} to improve the sensitivity with respect to standard reflection measurements,\cite{Giannetti2007} oscillates at the pseudo-SAW frequency $\nu_{0}$. The acoustic modulation's frequency $\nu_{0}$ and spectral width $\Gamma$ are extracted by Fourier transforming the signal. The device is exploitable provided the frequency shift $\Delta\nu$ due to mass loading\cite{Nardi2009} is resolvable.\\
\indent To understand when this condition is provided, we consider a maximum pump-probe time delay $\Delta t=10$~ns (corresponding to a 3 m long optical delay), thus limiting the minimum experimentally resolvable frequency shift $\Delta\nu_{ext}$ to $0.1$~GHz. The constraint $\Delta\nu_{ext}$ is $\textit{extrinsic}$, a longer optical delay line improving the resolution. Small values of the geometric ($h$, $f$) and mass loading ($\rho_{Al}$) factors limit the grating-assisted scattering of the pseudo-SAW into the bulk, therefore granting a long acoustic lifetime $\tau$. The pseudo-SAW lifetime for the free $\tau_{0}$, and loaded $\tau_{1}$, device depends on the acoustic properties of the composite system, and affects the minimum $\textit{intrinsically}$ detectable frequency shift $\Delta\nu_{int}$. We point out that the detectable $\Delta\nu\ge max\left\{\Delta\nu_{ext},\Delta\nu_{int}\right\}$. The subscripts $0$ and $1$ will refer from now on to the free and mass loaded device, respectively.\\
\indent In order to investigate the device's performance in terms of minimum detectable mass per unit of surface $\Delta m_{s}$ and sensitivity $S_{\nu}\equiv\displaystyle\lim_{\Delta m_{s}\to 0}\frac{1}{\nu_0}\frac{\Delta\nu}{\Delta m_{s}}$, calculations of $\left\{\textbf{u}_{i}(\textbf{r})\right\}$, $\textbf{u}(\textbf{r},t)$, and ultimately the absolute value of the projection coefficients $\left|c_i\right|=\left| \langle \textbf{u}_i|\textbf{u}\left(\textbf{r},t\right)\rangle \right|$, are performed on the proposed SPC configuration, adopting the theoretical framework outlined in Ref.~[\onlinecite{Nardi2011}].\\
\indent A critical issue is the evaluation of the frequency of the pseudo-SAW solution $\textbf{u}_{SAW}$, in relation to its strong surface confinement, calling for calculations beyond a perturbative approach.\cite{Nardi2009} We positively tested the reliability of the adopted framework in the hypersonic frequency range by benchmarking it against measurements reported by Siemens et al.\cite{Siemens2009} for Ni on sapphire SPCs working up to 47~GHz, as shown in Fig.~\ref{Fig1}. Excellent agreement is found between the experimental (blue empty circles) and calculated (red circles) pseudo-SAW dispersion curves. The deviation of the SPC's frequency dispersion curve from the pure Rayleigh wave solution, here reported for comparison (dashed black line), increases as the stripe's periodicity $p$ is reduced. An outlook of what can be expected beyond the 50~GHz range is provided for similar devices with $p=75$~nm and $50$~nm (green empty circle), the latter being the same periodicity adopted for the proposed mass sensor.\\
\indent The absolute value of the calculated projection coefficient $\left|c_i\right|$ profiles, reported in Fig.~\ref{Fig2}, highlight the eigenmode mainly contributing to the displacement field of the excited Al-SPC. To check the impact of the material choice, results for a similar device based on Ni stripes are also reported. The resonances are ascribed to pseudo-SAW solutions, by inspecting their SAW-likeness coefficient\cite{Nardi2009} and displacement field $\textbf{u}_{SAW}$. Line shape analysis of the $\left|c_i\right|$ profiles gives direct information on the pseudo-SAW lifetime, and consequently on $\Delta\nu_{int}$. The interaction between the discrete surface eigenmode and the continuum of bulk modes, induced in the composite system by the surface stress at the nanostripe/substrate interface, suggests fitting the $\left|c_i\right|$ data set with a Fano line shape profile:\cite{Nardi2011} 
\begin{equation}
F(\nu)= F_{o}+A\frac{\left(\left(\nu-\nu_{res}\right)+q\frac{\Gamma}{2}\right)^2}{\left(\nu-\nu_{res}\right)^{2}+\left(\frac{\Gamma}{2}\right)^{2}} \;,
\label{Fano}
\end{equation} 
where the profile index $q$ carries information on the configuration interaction, $\nu_{res}$ matches the pseudo-SAW device frequency $\nu_0$ ($\nu_1$), $F_{o}$ is the fit vertical offset, A is a constant, and $\Gamma$ is the line broadening parameter.\\
\begin{table}[t]
\begin{center}
\centering \caption{\label{Mass_loading} SPC mass sensors' performances. Al-SPC is the device proposed in the present work. Ni-SPC is the Ni-based version. Their operating frequency, and the line broadening parameter of the calculated projection profiles, are reported for different mass loading configurations.\\}
\begin{ruledtabular}
\begin{tabular}{crrr}
 &  & Al-SPC & Ni-SPC\\
\hline
Free & $\nu_0$ (GHz) & 127.78 & 125.23\\
 & $\Gamma_0$ (GHz) & 0.10 & 0.90\\
 & $\Delta\nu_{ext}$ (GHz) & 0.10 & 0.10\\
\hline
$\Delta m_s=10$ & $\nu_1$ (GHz) & 127.66 & 125.10\\
(ng/cm$^2$) & $\Delta\nu$ (GHz) & 0.12 & 0.13\\
 & $\Delta\nu_{int}$ (GHz) & 0.06 & 0.47\\
\hline
$\Delta m_s=50$ & $\nu_1$ (GHz) & 127.17 & 124.57\\
(ng/cm$^2$) & $\Delta\nu$ (GHz) & 0.61 & 0.66\\
 & $\Delta\nu_{int}$ (GHz) & 0.10 & 0.60\\
\end{tabular}
\end{ruledtabular}
\end{center}
\end{table}
\indent The performance of the mass sensor is assessed comparing the projection profiles for the free and loaded device, as seen in Fig.~\ref{Fig2}. To access the pseudo-SAW frequency shift $\Delta\nu=\nu_0-\nu_1$, we must resolve the resonance peaks for the free and loaded device, highlighted in the projection profiles of Fig.~\ref{Fig2}(b) and (c). The criterion we adopt is $\Delta\nu_{int}=\Gamma_{1}/2$.\cite{note3}. The profiles' line broadening and the devices' operating frequency are summarized in Table~\ref{Mass_loading}. Upon a mass loading of $\Delta m_{s}=50$~ng/cm$^2$, the Al-SPC resonance shifts downward by $\Delta\nu=0.61$~GHz, resulting in $\Delta\nu>max\left\{\Delta\nu_{ext},\Delta\nu_{int}\right\}$, well within the detectability range. On the other hand, under the same loading conditions the downshift of the Ni-SPC resonance by $\Delta\nu=0.66$~GHz results in $\Delta\nu\sim\Delta\nu_{int}>\Delta\nu_{ext}$. This value for $\Delta m_{s}$ is therefore the minimum detectable for the Ni-based device, the limitation being \textit{intrinsic}. The minimum detectable $\Delta m_{s}$ for Al-SPC is $10$~ng/cm$^2$. Calculations for this mass loading on both devices are reported in Fig.~\ref{Fig2}(c). The Al-SPC resonance shifts by $\Delta\nu=0.12$~GHz, resulting in $\Delta\nu\sim\Delta\nu_{ext}>\Delta\nu_{int}$, therefore the limitation being \textit{extrinsic}, as evidenced in the inset of Fig.~\ref{Fig2}(c). The pseudo-SAW frequency shift is not resolvable for the Ni-based device, the demonstration being $\Delta\nu=0.13$~GHz $<\Delta\nu_{int}$.\\
\begin{table}[t]
\begin{center}
\centering \caption{\label{Sensor_comparison} Comparison of mass sensors' sensitivities. QCM, SAW and FPW stand for, respectively, quartz crystal microbalance, typical surface acoustic wave-based and flexural plate wave-based sensors, see Ref.~[\onlinecite{Cheeke1999}].\\}
\begin{ruledtabular}
\begin{tabular}{lcc}
Device & $\nu_{0}$ & $S_{\nu_{0}}$\\
type & (GHz) & (cm$^{2}$/g)\\
\hline
Al-SPC & 127.78 & 95000\\
Ni-SPC & 125.23 & 105000\\
QCM & 6.0$\times$10$^{-3}$ & 14\\
SAW & 0.11 & 151\\
FPW & 2.6$\times$10$^{-3}$ & 951\\
\end{tabular}
\end{ruledtabular}
\end{center}
\end{table}
\indent In Table~\ref{Sensor_comparison}, we compare the performance of the outlined mass sensor. Despite similar sensitivities, the proposed Al-SPC mass sensor outperforms the corresponding Ni-based one in terms of minimum detectable $\Delta m_{s}$. Furthermore, our device's sensitivity is nearly three orders of magnitude higher than typical figures for standard IDT sensors, thus proving the validity of the proposed design. Viewed the outstanding improvement in sensitivity, the stringent geometric requirements and metal stripes material choice could be relaxed, the device design still being competitive with respect to current technology. Within this scheme, acoustic wave generation/detection and mass loading occur in the same active region, identified by the probe beam spot size. This feature allows for increased miniaturization with respect to IDT-based devices and enables operation with minute amounts of matter. The probe diameter can in principle be scaled to $\lambda_{probe}$. For instance, in the case of Al-SPC a probe beam focused to a 10~$\mu$m diameter is scattered by 200 metal stripes,\cite{note1} detecting a minimum deposited mass $\Delta m=8$~fg.\\
\indent In conclusion, a design for a photoacoustic mass sensor operating at 127 GHz is theoretically investigated, based on impulsive optical excitation of pseudo-SAWs in a SPC with 50 nm periodicity, and time-resolved EUV detection of the pseudo-SAW frequency shift upon mass loading of the device. The theoretical framework used for device engineering has been positively benchmarked against published data.\cite{Siemens2009} The device incorporates time-resolving capabilities, allows for increased miniaturization, enables operations with minute amounts of matter, and does not require a piezoelectric substrate, widening the range of exploitable materials with respect to standard mass sensors technology. The calculated device sensitivity outperforms standard electro-acoustical technology. In perspective, further increase in sensitivity may be achieved by exploiting SPC's pseudo-SAW long-living modes at the edge of the Brillouin zone ($k_{x}\le\pi/L$, $n$=0), to overcome the ultimate bottleneck standing in the pseudo-SAW linewidth $\Gamma$, i.e. the coupling between pseudo-SAWs and bulk modes.\cite{Maznev2008} This could be accomplished by extending the transient grating technique, applied to SPCs,\cite{Maznev2009} to the EUV frequency range.\cite{Tobey2006} The set of time-resolved optical techniques exploitable for mass sensing within the proposed device scheme might be expanded. For instance, asynchronous optical sampling (ASOPS) might enable infrared detection in a direct reflection geometry. The ASOPS technique,\cite{Bartels2007} avoiding all mechanical movement and with fast sampling rate, proved capable of detecting relative reflectivity variation in the $10^{-7}$ range. Appropriate tuning of the repetition rate would allow lowering the extrinsic frequency resolution to the 10 MHz range.\cite{Bruchhausen2011} The advantages gained in data acquisition rate might, as a matter of fact, compensate for the lower sensitivity as compared to the EUV diffraction signal detection.\\
\acknowledgments{We would like to thank Dr. Mark E. Siemens, Dr. Qing Li, Prof. Margaret M. Murnane and Prof. Henry C. Kapteyn for valuable discussions and for allowing us to reproduce their data from Ref.~[\onlinecite{Siemens2009}]. We also thank Prof. Fulvio Parmigiani for valuable discussions. The authors acknowledge partial financial support by MIUR-PRIN 2008 project.}
\begin{widetext}  
Copyright (2012) American Institute of Physics. This article may be downloaded for personal use only. Any other use requires prior permission of the author and the American Institute of Physics.
The article appeared in Appl. Phys. Lett. \emph{100}, 253106 (2012); doi: 10.1063/1.4729624 and may be found at http://dx.doi.org/10.1063/1.4729624
\end{widetext}

\begin{thebibliography}{18}
\bibitem[{not({\natexlab{b}})}]{note1}
\bibinfo{note}{The Al-SPC's first order diffraction efficiency, for
  $\lambda_{probe}=30$ nm, is $\eta=2\times10^{-4}$. Assuming a 1 mJ laser pulse
  entering the nonlinear fiber and an EUV conversion efficiency of 10$^{-6}$,
  the EUV energy per pulse is 1 nJ. For a 2 KHz repetition rate laser, the
  photon flux at the detector is therefore $\sim5.5\times10^{8}$ photon/s, well
  within range of EUV detectors capabilities.}
\bibitem[{not({\natexlab{c}})}]{note3}
\bibinfo{note}{The expression is obtained from the more general form
  $\Delta\nu_{int}=max\left\{\Gamma_{0},\Gamma_{1}\right\}/2$, for
  $\Gamma_1>\Gamma_0$, which is the case for SPC mass sensors.}
\end{thebibliography}
\end{document}